\documentclass[aps,floatfix,showpacs,prb]{revtex4}

\begin{document}
%%%%%%%%%%%%%%%%%%%%%%%%%%%%%%%%%%%%%%%%%%%%%%%%%%%%%%%%%%%%%%%%%%%%%%%%%%%%%%%%%%

\title{Interlayer phase coherence and composite fermions}

\author{Thierry Jolicoeur}
\affiliation{Universit\'e Paris-Saclay, CNRS, CEA, 
Institut de Physique Th\'eorique, France}

\date{January 25th, 2025}

\begin{abstract}
The fractional quantum Hall effect (FQHE) realized in two-dimensional electron systems
is explained by the emergent composite fermions (CF) out of ordinary electrons.
It is possible to write down explicit wavefunctions explaining many if not all
FQHE states. In bilayer systems there is a regime at integer filling of the lowest Landau level that displays a spontaneous breakdown of the U(1) relative phase between
the two layers. This can be seen as interlayer phase coherence (ILC) in terms of \textit{electrons}. Recent experiments in double layer samples of graphene
have revealed the appearance of many FQHE states unique to the bilayer case.
We discuss extensions of the CF idea in this situation as well as the possible existence of ILC of CFs.
\end{abstract}

\maketitle

\section{Introduction}
In the low-temperature regime quantum fluids may display physical properties
governed by spontaneous symmetry breaking. This is the case of superfluids
with the symmetry associated to the number of neutral particles. This also happens in superconductors with breakdown of the gauge symmetry of electrodynamics due to the condensation of Cooper pairs. Another set of
phenomenon is due to the development of strong topological order
known to happen in two-dimensional electrons gases in a magnetic field.
This is the realm of the fractional quantum Hall effect~\cite{Tsui1982,Laughlin} (FQHE). In this short review we will discuss the interplay of both phenomena
in the context of graphene bilayers. Indeed it has been possible recently
to fabricate devices by stacking monolayers of graphene separated
by a barrier of hexagonal boron nitride leading to an atomic scale
physical separation between the layers and essentially no tunneling.
Such devices have revealed a complex pattern of FQHE states and some
of them may be described by spontaneous interlayer phase coherence
of emerging entities~\cite{jainbook} called ``composite fermions''.
These composite fermions are in this case even more complex objects~\cite{CsathyJain}
than the ones appearing in single layer devices.

We will focus only on the approaches based on explicit wavefunctions
written in terms of electron coordinates that have been shown
to be extremely useful~\cite{jainbook,scarola,KamillaCF}.

\section{fractional quantum Hall effect}
To set the stage for the wavefunctions describing FQHE we first give
a short introduction to Landau level physics and the integer quantum Hall
effect which does not involve in a crucial way electron-electron interactions.
The emerging dressed electrons called ``composite fermions'' that describe the FQHE will
essentially be ruled by this regime. Numerous books are available, giving
all necessary details~\cite{jainbook,dassarma,heinonenbook,Prange}.

\subsection{Integer quantum Hall effect}
We consider a system of $N_e$ electrons confined to two-dimensional plane under a
perpendicular uniform magnetic field. This is a physical situation describing
some electron gases in semiconductors, heterostructures or quantum wells.
A simple description of this situation is given by the following Hamiltonian~:
\begin{equation}
\mathcal{H}=\frac{1}{2m^*}(\mathbf{p}+e\mathcal{A})^2,
 \label{defHam2D}
\end{equation}
where the vector potential can be taken in the so-called symmetric gauge
$\mathcal{A}=(\mathcal{B}\times\mathbf{r})/2$. The effective mass $m^*$
is a material dependent parameter. Solving for eigenvalues of the one-body Hamiltonian Eq.(\ref{defHam2D}) gives the Landau levels that have energies
$E_N=(N+1/2)\hbar\omega_c$ with $N$ a positive integer and $\omega_c=eB/m^*$
the cyclotron frequency.
There a macroscopic degeneracy in this situation~: for each value of $N$
there are exactly $N_\phi$ states that have this peculiar energy
where $N_\phi$ is the number of flux quanta through the system $N_\phi=B\times A/(h/e)$. The FQHE is strongest when all electrons lie in the lowest Landau level (LLL)
whose one-body eigenstates are given by the following formula~:
\begin{equation}
 \phi_M(\mathbf{r})=z^M \exp (-|z|^2/4\ell^2),
 \label{onebody}
\end{equation}
where we use the complex coordinate in the plane $z=x-iy$ and we have defined
the magnetic length $\ell=\sqrt{\hbar/(eB)}$. The power $M$ is a positive or zero integer (we omit the normalization).
With the special Landau level spectrum it is clear that even without considering interactions there are special cases when one has exact filling of an \textit{integer}
number of Landau levels, say $p$. With the formula for the degeneracy this happens when~:
\begin{equation}
 \nu = \frac{N_e}{N_\phi} = \frac{nh}{eB} = p,
 \label{fillfactor}
\end{equation}
where we have defined the filling factor $\nu$ which is the fraction of occupied states, and $n$ is the areal density of electrons. A naive thought is that
in this integer case one has a magnetic-field induced insulator due to the
presence of cyclotron gap $\hbar\omega_c$.
This gap lies between the highest occupied $p$th Landau level
and the lowest empty $p+1$th level at zero temperature.
This is the physical picture for the bulk sample. In a real
finite-size sample the Landau levels are bent when approaching the boundaries
and as a consequence the Landau levels will cross the Fermi energy
at some points that we may call
``Fermi points'' at the real-space edges of the system.
As a consequence there is electric conduction with appearance of a Hall voltage
$V_H$
perpendicular to an imposed current $I$ and its value is given by~:
\begin{equation}
 V_H = \frac{1}{p}\frac{h}{e^2}I= \frac{1}{\nu}\frac{h}{e^2}I,
 \label{HallRes}
\end{equation}
so that the Hall resistance defined by $R_H=V_H/I$ is quantized in units of
$h/e^2$ and $I$ is the current intensity in the sample. The presence of disorder in the sample leads to the appearance
of plateaus as as a function of the magnetic field where the Hall resistance stays
constant at the value Eq.(\ref{HallRes}). This phenomenon, the so-called integer
quantum Hall effect~\cite{Halperin1983}, does not involve in any crucial way the electron-electron interactions. 

The closed-shell argument above suggests that for fractional filling
no similar quantization should develop since partial filling of a flat band
enlarged by disorder may host a Fermi liquid albeit with a large density of
states. But striking experiments found evidence for the existence
of a fractional quantum Hall effect when the filling factor $\nu$ is
some rational fraction and the phenomenology of current transport
is the same as in the integer case. The original discovery involved
a plateau in the Hall resistance $R_H=3h/e^2$ corresponding to filling
factor $\nu=1/3$ in the LLL. To understand the fractional case one has to
dig deep into the many-body problem of interacting electrons in the LLL.
A generic many-body wavefunction is a sum of products of one-body states Eq.(\ref{onebody}) and thus can be written as a polynomial $\mathcal{P}$
in complex coordinates $z_i$ times a universal (state-independent) Gaussian factor~:
\begin{equation}
 \Psi(z_1,\dots,z_N)=\mathcal{P}(z_1,\dots,z_N) \,\,\exp (-\sum_i |z_i|^2/4\ell^2).
\end{equation}
In the spin-polarized case the polynomial $\mathcal{P}$ is antisymmetric to comply
with the Pauli principle. In the general case it is hard to guess what are the relevant polynomials $\mathcal{P}$ but in the realm of the FQHE the successfull approach
leads to explicit guess of the polynomials, the most famous case being the Laughlin wavefunction. We conclude this section by the giving the expression of $\mathcal{P}$ for the completely filled LLL. We first observe that the one-body states Eq.(\ref{onebody}) has a probability distribution which is a ring centered
at $z=0$ (because of the choice of gauge) whose radius grows with the exponent $M$.
So fully filled LLL means that we occupy all  one-body states from $M=0$ up to
a maximal value given by $M=N-1$ without any vacancy $z^0,z^1,z^2,\dots,z^{N-1}$.
This state is exactly
a Slater determinant~:
\begin{equation}
 \Psi_{\nu=1}=\det(z_i^{j-1})=\prod_{i<j}(z_i-z_j),\quad i,j=1,\dots,N,
 \label{VDM}
\end{equation}
where we have used the fact that the determinant is the so-called Vandermonde determinant.
From now on we will omit the Gaussian factor from the one-body states since
it is independent of the state under consideration.
If we draw the charge density of state $\Psi_{\nu=1}$ in real space we find
that it has the shape of a very flat pancake with uniform density in its interior
and going to zero very quickly at the edge with a characteristic length $\ell$.

\subsection{Laughlin wavefunction}
To explain the properties of the quantum state of electrons at $\nu=1/3$ Laughlin
proposed the following candidate wavefunction~:
\begin{equation}
 \Psi^{(3)}=\prod_{i<j}(z_i-z_j)^3 \exp (-\sum_i |z_i|^2/4\ell^2).
\end{equation}
If we compute the charge density by some means we find also a very flat pancake 
as in
state Eq.(\ref{VDM}) but with a mean density three times lower, corresponding
to a uniform electronic state with filling factor 1/3 (this would be $1/m$
if we were to put power $m$ to the Vandermonde factor).
This wavefunction has very good trial energy but is not an exact eigenstate of the Coulomb interacting electrons. So its theoretical status is not immediately clear.
Another ansatz was the idea of starting from the Slater determinant of $\nu=1$
and fill only one out of three orbitals~: $z^0,z^3,z^6,\dots$. Since this is still a determinant it can be easily computed giving $\prod_{i<j}(z_i^3-z_j^3)$.
This is not a good candidate for the FQHE state at $\nu=1/3$. While it has the correct filling factor by construction it has higher energy than the Laughlin state
because it lacks the big correlation hole around each electron due to the 
$(z_i-z_j)^3$ factor in the Laughlin wavefunction. Also this state called the Tao-Thouless state does not have a uniform density in real space and does not lead
to the correct phenomenology of the FQHE at $\nu=1/3$, contrary to the Laughlin
state.

To clarify the status of the Laughlin state we now show to solve by elementary
means any two-body problem in the LLL. The interacting Hamiltonian is~:
\begin{equation}
 \mathcal{H}_2=\frac{1}{2m^*}(\mathbf{p}_1+e\mathcal{A}_{r_1})^2+
 \frac{1}{2m^*}(\mathbf{p}_2+e\mathcal{A}_{r_2})^2 + V(r_1-r_2),
\end{equation}
where the positions of the electrons are $r_1,r_2$ and $V$ is the interaction potential. In the context of the FQHE it is given by the Coulomb interaction
$V(r)=e^2/r$. We now introduce the center of mass and relative coordinates
$R=(r_1+r_2)/2$, $r=r_1-r_2$. The fact that the vector potential is linear
in the coordinates leads to a simplification:
\begin{equation}
 \mathcal{H}_2=\frac{1}{2M}(\mathcal{P}+2e\mathcal{A}_{R})^2+
 \frac{1}{2\mu}(\mathbf{p}_r+(e/2)\mathcal{A}_{r})^2 + V(r),
\end{equation}
so the center of mass as well as the relative particle will live in their own
separate Landau levels. If we focus on the relative particle we can take as a basis
the one-body states of Eq.(\ref{onebody}) that are eigenstates of the angular momentum. If we consider a potential $V(r)$ which is rotationally invariant
it will be automatically diagonal in such a basis. The eigenenergies
are thus given by $V_m=\langle\phi_m|V(r)|\phi_m\rangle$. The numbers $V_m$
are often called pseudopotentials and are given by a simple integral.
In the Coulomb case the pseudopotentials decrease as $1/\sqrt{m}$. For spinless
electrons only the odd values of $m$ matter due to wavefunction antisymmetry.
As a consequence, the full many-body interacting Hamiltonian can be written as~:
\begin{equation}
 \mathcal{H}=\sum_{i<j}\, \sum_{m=0}^\infty \, V_m\,
 \, \mathcal{P}^{(m)}_{ij}
 \label{generic}
\end{equation}
where we have defined $\mathcal{P}^{(m)}_{ij}$ the projector onto the state
of relative angular momentum $m$ for the pair $i,j$ of particles.

If now we retain only the $m=1$ projector in the Hamiltonian Eq.(\ref{generic})
then we note that the Laughlin state is an exact zero-energy eigenstate.
Indeed since for all pairs of particles there is an overall factor $z^3$
with $z$ the relative particle of pair $i,j$ it means that it has no weight
onto $m=1$. For the experimentally relevant case of Coulomb interaction
the $V_{m=1}$ pseudopotential is indeed the strongest which gives some weight
to the relevance of the Laughlin state beyond the fact of having a good energy.
The parametrization of Eq.(\ref{generic}) suggests a way to interpolate between the Coulomb case and the hard-core limit involving only the projector $\mathcal{P}_1$ by varying the set of discrete
pseudopotentials $V_m$. This idea was successfully implemented by F. D. M. Haldane (see his contribution in book~\cite{Prange})
showing that the Laughlin state is smoothly connected to the Coulomb ground state
and thus these two states share the same physics.

As a polynomial in the electronic coordinates the Laughlin ansatz gives
us a state which is already fully factorized. 
This is a mathematical statement concerning the polynomial appearing in the
many-body wavefunction. It does not mean that the quantum state is factorizable,
the Laughlin state is \textit{not} a Slater determinant.
Indeed it is known to be a case of strong entanglement, a property which
does not exist in Slater determinant states.
The zeros of this polynomial
are located exactly at the positions of the electrons, a very special feature which is not true in general. 
Let us rewrite the Laughlin polynomial in the following way~:
\begin{equation}
 \prod_{i<j}(z_i-z_j)^3=\prod_{i<j}(z_i-z_j)\times \prod_{i< j}(z_i-z_j)^2
 =\det(z_i^{j-1}) \times \prod_{i< j}(z_i-z_j)^2.
\end{equation}
Here we have explcitly factored out one power of the Vandermonde determinant
Eq.(\ref{VDM}). We observe that 
if we pick  one electron 
and drag it around the closest neighbor we see that the phase of the
wavefunction changes by $6\pi$ while antisymmetry due to the Pauli
principle requires only a $2\pi$ turn. We interpret this property
by saying that there are exactly two vortices bound to each electron
in this state of matter. The vortex attachment is directly due
to the extra factor $\prod_{i< j}(z_i-z_j)^2$. While this reveals
a fundamental property of the FQHE state, it does not lead immediately
to a recipe to construct other candidate wavefunctions for filling
factors beyond $\nu=1/3$. We now show in the next section
how another rewriting of the Laughlin polynomial leads naturally
to the so-called composite fermion construction.

\subsection{Composite fermion wavefunctions}
\label{JCF}
Experiments revealed also the appearance of the FQHE for fractions other than 1/3.
The most prominent set of such states appears for filling factors
$\nu=p/(2p\pm 1)$ with $p$ integer, asking for an explanation in terms of wavefunctions beyond the 
Laughlin state. One such state is given by the so-called composite fermion
construction that we describe now. Much of our intuition of electron systems is based on Slater determinants and occupied/empty orbitals so the first step is to rephrase the successfull Laughlin state in this language. We write the 
correlation factor of the Laughlin state in the following way:
\begin{equation}
 \prod_{i<j}(z_i-z_j)^3=\prod_{i<j}(z_i-z_j)\times \prod_{i\neq j}(z_i-z_j)
 =\det(z_i^{j-1}) \times \prod_{i\neq j}(z_i-z_j).
\end{equation}
Distributing the last among the columns of the determinant we arrive at the following identity~:
\begin{equation}
 \prod_{i<j}(z_i-z_j)^3
 =\det(z_i^{j-1} \prod_{ k\neq i}(z_i-z_k)).
\end{equation}
This can be interpreted as a Slater determinant for a filled Landau level
provided one replaces the one-body wavefunctions by correlated one-body
wavefunctions~:
\begin{equation}
 z^m \rightarrow z^m \prod_j (z-z_j),
 \label{Jastrow}
\end{equation}
where the product over $j$ is over all other particles.
The extra correlation factor is called a Jastrow factor in many-body physics.
If we decide to adopt such correlated orbitals instead of the one-body states
we can view the Laughlin state at $\nu=1/3$ as a completely filled Landau level
with $\nu^*=1$. We guess then that excited states may involve higher-lying Landau levels provided one uses modified orbitals including the correlation factor
Eq.(\ref{Jastrow}). In Landau levels other than the lowest the one-body
eigenstates involve the complex conjugate $z^*$ in addition of the complex coordinate $z$. Let us call $\phi_{N,m}(z,z^*)$ such a state in the $N$th Landau level.
We add Laughlin-style correlations by making the product as above~:
\begin{equation}
 \phi_{N,m}(z,z^*)\rightarrow \phi_{N,m}(z,z^*)  \prod_j (z-z_j).
\end{equation}
Such a state however does not live in the LLL due to the appearance
of $z^*$ factors. Since there is overwhelming evidence for FQHE in the LLL
only we want to project such states into the LLL and then play the game
of making Slater determinants. The operation of projection onto the LLL
amounts to putting all $z^*$  to the left-hand side of the formula and
next replacing them by the operator $2\partial/\partial z$. The result is then
entirely in the LLL. In the $N=1$ Landau level this manipulation amounts
to the substitution~:
\begin{equation}
  z^m \rightarrow z^m \frac{\partial}{\partial z}\prod_j (z-z_j),
\end{equation}
and going to even higher Landau levels simply add more derivatives acting onto
the correlation factor. Since the correlation factors map $\nu=1/3$ onto
$\nu^*=1$ one can say that the composite fermions feel a reduced magnetic
field $B^*=B-2n\phi_0$ by using the definition Eq.(\ref{fillfactor}) of the filling factor. This immediately suggests that there will be an IQHE for the
composite fermions when $\nu^*=p$ with $p$ filled Landau levels of the CFs.
This translates in filling factors for electrons as $\nu=p/(2p+1)$.
One can then write down Slater determinants made of correlated orbitals
to describe such states. This procedure gives us explicit trial wavefunctions
whose energies can be computed by a simple Metropolis sampling. Many if not all
properties extracted from these CF wavefunctions are in excellent agreement
with the known experimental data and also in agreement with numbers
obtained from exact diagonalization of systems with a (very) small number
of electrons. The series of FQHE states at $\nu=p/(2p+1)$ is prominent in high-quality samples and is called the Jain series of states. It is observed 
from $p=1$ (the Laughlin state) up to at least $p=10$. It is important to note
that these states have \textit{no} variational parameters and nevertheless offer
a very accurate description of FQHE states (for detailed comparisons see e.g. \cite{jainbook}). The heuristic mapping $B\rightarrow B^*$ also correctly predict that when $B^*=0$ the CFs form some kind of Fermi sea which is gapless
as is observed in electron gases at $\nu=1/2$.

With the notion of effective magnetic field $B^*$ we note that it may be negative,
leading to states with $\nu^*=-p$ hence $\nu=p/(2p-1)$ as observed in experiments. It is also easy to generalize the CF construction
to fractions descending from the parent state $\nu=1/5$. Indeed the Laughlin state
can accommodate any odd power of the Jastrow factor for spin polarized fermions
(even power for bosons). The same line of reasoning leads to series of states
with $\nu=p/(4p+1)$ and $\nu=p/(4p-1)$, again many of such states
are observed in nature. When the filling factors becomes low the FQHE
are competing with a crystal state made out of electrons, called the Wigner crystal which does not have the same striking properties as the FQHE states.

The CF construction also gives a very simple picture of excited states.
Since the CF are filling an integer of pseudo LLs a first type of excitations
consists of promoting a CF from the topmost filled pseudo LL to the lowest
empty pseudo LL. This is a neutral excitation with no change of the number 
of electrons or the number of flux quanta. One may expect that such an
excited
state has an energy cost given by the effective cyclotron energy for the CFs.
There are also charged excited states obtained by making a hole in the topmost filled pseudo LL~: such a state is called a quasihole. By reducing the applied magnetic flux one can also
create a situation with only one electron promoted to the next pseudo LL. This is then
the quasielectron state.

In the CF picture one still has to explain why the Hall resistance
is quantized as $R_{xy}=(1/\nu)(h/e^2)$ with the $\nu$ the \textit{electron}
filling while there are $p$ filled pseudo Landau levels of CF in the
series $\nu=p/(2p+1)$. The explanation is that while the CF contains
a charged electron it also binds two vortices. 
Indeed in the construction of the CF states the Jastrow factor
squared is always present and in line with the case of the Laughlin state
we interpret this factor by saying that there are two vortices
bound to each eletron.
The CF vortex carries
two units of flux $\phi_0$ and if such a vortex crosses a Hall bar
it will induce a voltage drop $e=-d\Phi/dt=2(h/e)(I/e)$ where $I/e$ is
the number of CF per unit time crossing the Hall bar. So there is an
additional contribution to the Hall voltage~:
$V_H=(1/p+2)h/e^2 I$. This completes the explanation of transport
phenomenology in the FQHE regime which is mapped onto that of the IQHE.

Finally we mention that the composite fermions may undergo a pairing instability. This was proposed by Moore and Read~\cite{MR} who introduced
yet another intriguing explicit wavefunction called the Pfaffian~:
\begin{equation}
 \Psi_{Pf}=\mathrm{Pf}\left(\frac{1}{z_i-z_j}\right)
 \prod_{i<j}(z_i-z_j)^2,
 \label{MR}
\end{equation}
where the symbol Pf stands for the Pfaffian of the square matrix
$1/(z_i-z_j)$. If we compute the determinant of an antisymmetric matrix
we find that it is the square of a polynomial of the matrix elements.
This polynomial is called the Pfaffian of the matrix. It appears when one projects
a paired state like the BCS wavefunction onto a state with fixed number
of particles. So the appearance of this peculiar  factor is indicative of the paired nature of the Pfaffian state. Indeed the Pfaffian state
is a $p$-wave paired state of composite fermions. The filling factor
of the state can be easily computed and is $\nu=1/2$.
This state is not an exact eigenstate of the Coulomb interaction and is
a trial state competing with the Fermi sea of composite fermions that exists at the same filling factor. In the LLL in semiconductors the Fermi
sea has lower energy and the $\nu=1/2$ state is compressible
but it may be that this state becomes the ground state in the next Landau level where the effective Coulomb interaction is different
from the LLL so it is a candidate for the FQHE
state observed for $\nu=2+1/2=5/2$ in some very clean devices.
This state supports excitations with non-Abelian statistics which
are very interesting quasiparticles. We will not discuss in more details
this important FQHE topic.

\section{Interlayer phase coherence}
\subsection{the role of spin}
We now discuss the modifications of previous ideas when we consider the spin
degree of freedom of charge carriers.
The first remark is that the $\nu=1$ state should be written as~:
\begin{equation}
 \Psi^{(1)}=\prod_{i<j}(z_i-z_j)
 |\uparrow\dots\uparrow\rangle
 \label{spinnu1}
\end{equation}
since by construction it is fully polarized. More general states
will not be a simple product of a spin part times an orbital part.
If we consider the many-body problem of electrons interacting by the Coulomb
potential we note that it has full $SU(2)$ spin rotation symmetry.
This rotation symmetry will be broken down to $U(1)$ by the Zeeman coupling
of the external field to the total spin. The state (\ref{spinnu1})
has a spin projection onto the $z$ axis equal to $S^z_{tot}=+N/2$
and by rotational symmetry of the Hamiltonian it is also member of a multiplet
of total spin $S_{tot}=S^z_{tot}$. This multiplet is exactly degenerate without
Zeeman effect.
\begin{equation}
 \Psi^{(1)}=\prod_{k}c^\dagger_{k\uparrow}|0\rangle
 \label{spinnu1-2nd}
\end{equation}
The $S^z_{tot}=0$ member of this spin multiplet is obtained
by acting repeatedly with the spin lowering operator~:
\begin{equation}
 |S^z=0\rangle =(S^-_{tot})^{N/2}\,\,\prod_{k}c^\dagger_{k\uparrow}|0\rangle .
 \label{spinzero}
\end{equation}
In this formula the spin operator acts only on the spin degrees of freedom but
does not change the orbital part of the state. If we write the state in first quantization we decide to call $z_i$ the coordinates of $\uparrow$ spins
and $w_k$ of $\downarrow$ spins and the orbital part of the $S^z_{tot}=0$
state is then~:
\begin{equation}
 \Psi^{(1)}=\prod_{i<j}(z_i-z_j)\prod_{k<l}(w_k-w_l)
 \prod_{i,k}(z_i-w_k) ,
\end{equation}
The full wavefunction with the spin part is the antisymmetrized product
of this orbital factor and of the zero spin state 
$|\uparrow\dots\uparrow\downarrow\dots\downarrow\rangle$.
Halperin has proposed a generalization of the Laughlin wavefunction appropriate
to states involving spin~:
\begin{equation}
 \Psi^{(mmn)}=\prod_{i<j}(z_i-z_j)^m\prod_{k<l}(w_k-w_l)^m
 \prod_{i,k}(z_i-w_k)^n ,
 \label{Hal}
\end{equation}
Again this only the orbital part - it has to be supplemented by the spin part 
and antisymmetrized. Evaluation of spin-independent observables like the 
Coulomb energy only involve the orbital part so we simply omit the spin part for clarity. The filling factor of the trial state $mmn$ has to be computed
and is found to be $\nu=2/(m+n)$ where the filling factor refers to the
total filling including both species. 
In general it is not an eigenstate
of total spin. Special cases include $m=n+1$ which is a singlet $S_{tot}=0$
and $m=n$ states that are ferromagnetic states as the state Eq.(\ref{spinnu1}).
For example the $(332)$ Halperin state describes the singlet state $\nu=2/5$
that is a FQHE state appearing in samples with small Zeeman effect.

We note that at a given filling factor there are several Halperin state
that are competing. For example at $\nu_{tot}=1/3$ one may construct
the $(333)$ state which has Laughlin correlations irrespective
of the particle index and one can also build the $(551)$ state which may
become relevant when repulsion is weaker between the two components.

\subsection{quantum Hall bilayers}

There are several physical situations where the electrons have a pseudospin
index. This happens notably in many-valleys semiconductors. Monolayer graphene
has two valleys and thus an extra pseudospin with two values in addition to the
real spin. Some semiconductors like Si have up to six valleys. It may
happens that the Coulomb interaction is independent of these extra degrees of
freedom. This is approximately the case of monolayer graphene.
We discuss now the case of engineered systems where two spatially separated
layers of two-dimensional electron gases are close enough so that there are sizable Coulomb interactions. The layer index is then a pseudospin
and the Coulomb interaction is different inside a given layer and between layers~:
\begin{equation}
 V_{\uparrow\uparrow}=V_{\downarrow\downarrow}=e^2/r, \quad
  V_{\uparrow\downarrow}=e^2/\sqrt{r^2+d^2},
\end{equation}
where $d$ is the distance between layers. With this interaction the problem
does not have the full $SU(2)$ rotation symmetry in pseudospin space
but only the $U(1)$ rotation around the $z$ axis which is the conservation
law of the difference of particle numbers in the two layers. Let us concentrate
on the case $\nu=1$ first. In the limit $d\rightarrow 0$ we are back
to the symmetric situation discusses in the previous section and the ground state
is a ferromagnetic spin multiplet. If we now tune $d$ small the members
of the multiplet will no longer be degenerate. Since electron-electron interactions are weaker when they are in separate layers it means that
the $S^z_{tot}=0$ state will have lower energy than all other states
in the multiplet.

\subsection{symmetry breaking and the phase}

If we increase the number of electrons at fixed interlayer distance $d$
we discover that the multiplet of states $S^z_{tot}=-N/2,\dots,+N/2$
becomes degenerate as $N\rightarrow\infty$. This emergent degeneracy
is the hallmark of broken symmetry. Since these states differ by the transfer of electron between layer it is the XY symmetry associated to the relative
phase between layers which is broken. As in situations involving broken symmetry
the physics become transparent once we use a function which breaks explicitly the symmetry. A simple choice is~:
\begin{equation}
 \Psi_x=\prod_k c^\dagger_{k, x}|0\rangle
 =\prod_k (c^\dagger_{k, \uparrow}+c^\dagger_{k, \downarrow})|0\rangle ,
\end{equation}
where $k$ labels the one-body states of the LLL.
This state has $\langle S^z_{tot}\rangle=0$ even though it is not 
an eigenstate of $S^z_{tot}$. Expansion of the second formula reveals
that it has weight over all members of the multiplet. This state
is the analog of the BCS wavefunction for broken particle number. Its generalization
to an arbitrary relative phase  $\phi$ between layers is then~:
\begin{equation}
 \Psi(\phi)=\prod_k (c^\dagger_{k, \uparrow}+\mathrm{e}^{i\phi}
 c^\dagger_{k, \downarrow})|0\rangle
 \label{psiphi}
\end{equation}
This state gives equal weight to the two layers and this broken symmetry
is aptly called interlayer phase coherence.
In the magnetic language we are dealing with XY symmetry breaking
and the relative phase $\phi$ is the XY order parameter.
This phenomenon has been observed in semiconductor devices~\cite{Eisenstein} for total
filling factor unity and $d \lesssim\ell$. If the layer separation
is too large then we have two essentially decoupled layers each hosting
a $\nu=1/2$ CF Fermi sea. There is at least one transition between
the fully decoupled regime at large separation and the ILC phase at small separation. As expected in a system with a phase associated with symmetry
breaking there should supercurrents in states with a gradient in space of this
phase. However since the phase corresponds to the relative phase between the layers it implies that the supercurrent consists of opposite flows
of charge carriers in opposite layers so with zero total current.

Let us now perform a particle-hole transformation on only one spin species
of the state Eq.(\ref{psiphi}) and call 
$d_{k,\uparrow},d_{k,\uparrow}^\dagger$ the associated creation/annihilation operators. Then the state with definite phase Eq.(\ref{psiphi}) 
can be written as~:
\begin{equation}
  \Psi(\phi)=\prod_k (1+\mathrm{e}^{i\phi}
c^\dagger_{k, \downarrow}d_{k, \uparrow})|\bar{0}\rangle
 \label{excphi}
\end{equation}
where we have defined the new vacuum $|\bar{0}\rangle=\prod_k c^\dagger_{k, \uparrow}|0\rangle$. This new writing show that the ILC state
at total filling factor $\nu=1$ can be aptly called an exciton condensate
where the two members of the exciton pairs reside in different layers.

There are two configurations of currents that can be used to reveal
the ILC. The first one is the drag configuration in which a current
is imposed only one layer (the ``drive'' layer) and may then measure the Hall voltage across the drive layer giving then a measurement of $R_{xy}^{drive}$ and measure
the Hall voltage across the other layer with no drive current (the ``drag'' layer) giving access to $R_{xy}^{drag}$. If we consider the Halperin (111)
wavefunction we note that driving an electron in one layer is accompanied
by one vortex in the other layer so ILC has the special value~:
\begin{equation}
 R_{xy}^{drive} = R_{xy}^{drag} = \frac{h}{e^2}
\end{equation}
The other configuration is the counterflow set-up in which the current of the drive layer is injected at the end of the Hall bar backwards in the top layer
inducing a regime in which the currents flow in opposite directions in the two layers. If now we tune the magnetic field to reach $\nu_{tot}=1$
electrons and holes are locked together as in Eq.(\ref{psiphi}) and
thus form a neutral entity that does not feel the Lorenz force. So
the Hall voltage in both layers goes to zero right at $\nu_{tot}=1$
a striking evidence of ILC~\cite{Eisenstein}.

\section{Graphene bilayers in the modern era}

It has been feasible recently to build devices with stacking of two
atomically thin monolayers of graphene. Such systems can reach
a regime where $d/\ell\approx 0.1$ which was out of reach of previous
set-ups. Combined with high electronic mobility experiments have revealed
numerous FQHE states~\cite{NatPhysKim,NatPhysDean} that do not match the fillings of one-component
states discussed in section (\ref{JCF}). One can first ask whether
there is an extension of the CF wavefunctions giving trial wavefunctions for these states. Another more intriguing question is whether one can observe ILC
involving as basic building blocks CF instead of electrons.

\subsection{${}^2_1CF$ wavefunctions}

The Halperin family of states Eq.(\ref{Hal}) suggests a simple way to construct CF states~\cite{scarola}.
We take a CF state in each layer, make a product and add a correlation
factor between the two layers~:
\begin{equation}
 \Psi=\Psi_{\bar\nu}(z_i)\times\Psi_{\bar\nu}(w_k)\times\prod_{i,k}(z_i-w_k)^m .
\end{equation}
Now the exponent $m$ can be even or odd since there is no restriction
from Fermi statistics. One may expect that the relevant values of $m$ 
are smaller for larger separations between layers.
This state has  filling factors $\nu_{1,2}$ for each layer~:
\begin{equation}
 \frac{1}{\nu_{1,2}}=\frac{1}{\bar\nu}+m
 = \frac{1}{\nu^*}+2p+m
\end{equation}
The CF that are formed in such states are called ${}^{2p}_mCF$
where there are $2p$ vortices attached to the electron in the same layer 
and $m$ interlayer vortices, according to the interpretation of Jastrow
factor in the CF language.

By increasing the distance $d$ between the layers we expect a weakening
of correlations and transitions~\cite{scarola} between competing states at fixed filling factor. For example at $\nu_{tot}=1/3$ one may have a ground state
described by the $(333)$ state at small separations which is a state
of ${}^{2}_3CF$. For larger separation we construct the $(551)$ at the same filling which is now a state of ${}^{4}_1CF$. For very large separations there should should no interlayer repulsion and so each layer should
form a Fermi sea of ${}^{6}CF$
(note that at such a low filling factor $\nu=1/6$ a Wigner crystal may form
instead of a FQHE state).
By considering compressible states
one may enlarge the familty of trial states. For example there is
the Fermi sea of ${}^{4}_2CF$ which is also competing at $\nu_{tot}=1/3$
that would lead to a compressible intermediate phase sandwiched
between $(333)$ and $(551)$ states if we guess that the ordering of phase
follow the power of the interlayer repulsion factor.
At filling factor $\nu_{tot}=1/2$ one may consider the Halperin
state $(331)$ which may be realized in GaAs/GaAlAs devices~\cite{Suen92,Eisenstein12}.

By similar reasoning we expect that for $\nu_{tot}=2/5$
by increasing the separation we find the $(332)$ state which is a
${}^{2}_2CF$, then a Fermi sea of ${}^{4}_1CF$, then the state $(550)$
which is a product of two Laughlin states at filling 1/5,
an incompressible state of ${}^{4}CF$.

It is also plausible~\cite{faugno} that other non-FQHE state enter the competition
like crystal states of electrons or of composite fermions.
Only partial theoretical analysis of this complicated situtation
is available.

\subsection{observed states}
By analogy with monolayer electronic gases one expects to find
a prominent series of FQHE states involving ${}^2_1CF$~\cite{scarola}
which means $p=1$ and $m=1$ hence $\nu_{1,2}=n/(3n+1)$ where $n$ is the
number of filled pseudo-Landau levels of CFs. Indeed the states
$\nu_{1,2}=1/4 \,(n=1), 2/7 \,(n=2), 3/10 \,(n=3)$ are observed as 
incompressible states. The negative-flux  series is also observed for 
$\nu_{1,2}=1/2 \,(n=-1), 2/5 \,(n=-2), 3/8 \,(n=-3)$.
These two series of states should converge to a compressible state
for $\nu_{1,2}=1/3$.
Some of these fractions have simple wavefunctions~: this is the case of the
$1/4$ state which involves two Laughlin states with one Jastrow power 
between the layers. The wavefunction is the $(331)$ state in the Halperin
family Eq.(\ref{Hal}).
But these series of states do not exhaust the observations made in graphene bilayers. 

Some additional fractions beyond the principal $n/(3n\pm1)$ series
include states with fractional $n$ and odd denominator such as 
$\nu_{1,2}=3/14$ corresponding to $n=3/5$ or $\nu_{1,2}=2/9$
with $n=2/3$. It may be that such corresponds to FQHE states
of composite fermions as observed in one-component systems
for fractions like $\nu=4/11$. They may fall in the general hierarchical
scheme of FQHE.

Another set of states appear for $\nu_{1,2}=1/3$ and $2/3$. These states 
do not manifest any drag Hall resistance so they are decoupled
FQHE states likely $(330)$ for $1/3$ and its particle-hole partner for 
$2/3$.
The state observed at $\nu_{1,2}=1/6$ may be a Halperin state
$(333)$ provided interlayer correlations are strong enough
which is not completely consistent with uncorrelated states nearby
at 1/3. As we observed in the case of multicomponent states
at a given filling factor there are several competing states
when we vary the strength of interlayer coupling.

Finally there is evidence for states with half-integer filling
of ${}^2_1CF$. This includes $\nu_{1,2}=1/5\, (n=1/2), 3/11 \,(n=3/2),
5/17 \,(n=5/2)$ and with negative flux $\nu_{1,2}=3/7 \,(n=-3/2), 
5/3 \,(n=-5/2)$.
If we take into account the ILC phenomenology as well as the CF construction
it is natural to conjecture that these states may display ILC of ${}^2_1CF$.

It is now a theoretical challenge to write down explicit wavefunctions
capturing these states that combine the formation of emerging quasiparticles (the CFs) and their condensation in a broken symmetry state.

\section{conclusions}
Recent experiments have given evidence for a fascinating interplay
of topological order with the creation of a new type of composite
fermions, the ${}^2_1CF$ entities. These CF quasiparticles may
form Landau levels leading to incompressible states that are
a generalization of the series of FQHE states already observed
in single-layer one-component systems. However these experiments also revealed
the appearence of other FQHE states whose description may involve
interlayer phase coherence of composite fermions. Such states
await detailed theoretical explanation. The technological advances in
the manipulation of layered structures has opened a whole new field of investigation of correlated quantum states and the transitions between them.

Of course there are many more examples of multicomponent quantum Hall systems
that we have not addressed in this short review.
The experiments we have briefly discussed in this review~\cite{NatPhysDean,NatPhysKim}
have a thin insulating barrier of hexagonal boron nitride between two graphene monolayers. The barrier is thin
enough to allow strong Coulomb interactions between the two electron gases
but nevertheless thick enough to suppress tunneling.
It is feasible to fabricate samples with nonzero tunneling. This leads
a new set of physical properties. Notably the spin texture of charge excitations
is changed and can be manipulated by tilting the applied magnetic field
away from the direction perpendicular to the layers.
The physics of this situation has been investigated in samples involving
two semiconductor quantum wells coupled by tunnel effect. Each one-body
quantum state in a well is then combined with its partner in the opposite well
giving rise to eigenstates that are symmetric-antisymmetric (SAS) doublets with
tunnel energy splitting $\Delta_{SAS}$. It is thus feasible to investigate
such a system as a function of the applied magnetic field and $\Delta_{SAS}$.
Such experiments have revealed the competition between one-component and two-component
quantum Hall states as well as between single-layer and bilayer Wigner
crystals~\cite{ManoharanShayegan,ManoharanSST}.

In the graphene world one can also study pure graphene bilayers where there is
chemical bonding between two monolayers. The case of Bernal stacking (AB) has been
investigated in detail in the quantum Hall regime. This is a very special case
since the central Landau level has an orbital degeneracy between levels with $N=0$
and $N=1$ character where $N$ is the Landau level index. This is in addition 
to the spin and valley degeneracies which are also present as in the case of monolayer
graphene. The central level
orbital degeneracy can be adjusted by applying an electric bias between the two layers
leading to a tunable quantum Hall system which can interpolate
between $N=0$ and $N=1$ FQHE physics. One expects that some if not all FQHE
states can be described by appropriately generalized CF wavefunctions.
Recent investigations~\cite{JunZhu} have revealed many FQHE states with \textit{even}
denominators that are outside the scope of CF wavefunctions but likely
to belong to the Pfaffian family of FQHE states or its particle-hole
partner dubbed the ``AntiPfaffian''.

It is likely that progress in sample creation or fabrication will lead to more
insights into these remarkable states of matter and more guidance for in-depth
theoretical studies.

\section{Acknowledgments}
I thank Carlos S\'a de Melo and Yvan Castin for giving me the opportunity
to present this review at the workshop ``Open questions in the quantum many-body problem''
held in Paris in July 2024 at Institut Henri Poincar\'e.

% The next command determines the bibliography style. Please do not
% change this.

%  This inserts the bib file
\bibliography{IHP}

\end{document}